\documentclass[12pt,preprint]{aastex}
\newcommand{\lsim}{\, \lower2truept\hbox{${< \atop\hbox{\raise4truept\hbox{$\sim$}}}$}\,}
\newcommand{\gsim}{\, \lower2truept\hbox{${> \atop\hbox{\raise4truept\hbox{$\sim$}}}$}\,}

\newcommand{\puncspace}{\ifmmode\,\else{\ifcat.\C{\if.\C\else\if,\C\else\if?\C\else%
\if:\C\else\if;\C\else\if-\C\else\if)\C\else\if/\C\else\if]\C\else\if'\C%
\else\space\fi\fi\fi\fi\fi\fi\fi\fi\fi\fi}%
\else\if\empty\C\else\if\space\C\else\space\fi\fi\fi}\fi}
\newcommand{\SP}{\let\\=\empty\futurelet\C\puncspace}

\shorttitle{UV and IR galaxies at z=0.6} 

\shortauthors{Xu et al.}

\begin{document}

\title{IR and UV Galaxies at z=0.6 ---
Evolution of Dust Attenuation and Stellar Mass as Revealed by SWIRE and GALEX}

\author{C. Kevin Xu \altaffilmark{1,5}, 
David Shupe\altaffilmark{6},
Veronique Buat\altaffilmark{9}, 
Michael Rowan-Robinson\altaffilmark{3}, 
Thomas Babbedge\altaffilmark{3},  
Jorge Iglesias-P\'{a}ramo\altaffilmark{10},
Tsutomu T. Takeuchi\altaffilmark{11},
Tom A. Barlow\altaffilmark{1},
Tim Conrow\altaffilmark{1},
Fan Fang\altaffilmark{6},
Karl Forster\altaffilmark{1},
Peter G. Friedman\altaffilmark{1},
Eduardo Gonzales-Solares\altaffilmark{8},
Carol Lonsdale\altaffilmark{5},
D. Christopher Martin\altaffilmark{1},
Patrick Morrissey\altaffilmark{1},
Susan G. Neff\altaffilmark{4},
David Schiminovich\altaffilmark{1},
Mark Seibert\altaffilmark{1},
Todd Small\altaffilmark{1},
Gene Smith\altaffilmark{7},
Jason Surace\altaffilmark{6},
Ted K. Wyder\altaffilmark{1}
}

\altaffiltext{1}{California Institute of Technology, MC 405-47, 1200 East
California Boulevard, Pasadena, CA 91125}
\altaffiltext{2}{Laboratoire d'Astrophysique de Marseille, BP 8, Traverse
du Siphon, 13376 Marseille Cedex 12, France}
\altaffiltext{3}{Astrophysics Group, 
Blackett Laboratory, Imperial College of Science 
Technology and Medicine, Prince Consort Road, London SW7 2BZ, UK}
\altaffiltext{4}{Laboratory for Astronomy and Solar Physics, NASA Goddard
Space Flight Center, Greenbelt, MD 20771}
\altaffiltext{5}{Infrared Processing and Analysis Center, 
California Institute of Technology 100-22, Pasadena, CA 91125}
\altaffiltext{6}{Spitzer Science Center, California Institute of Technology, Mail Stop 220-6, Pasadena, CA 91125}
\altaffiltext{7}{Center for Astrophysics and Space Sciences, 
University of California, San Diego, La Jolla, CA 92093-0424}
\altaffiltext{8}{Institute of Astronomy, Madingley Road, Cambridge CB3 0HA, UK}
\altaffiltext{9}{
Observatoire Astronomique Marseille Provence, Laboratoire 
d'Astrophysique de Marseille, 13012 Marseille, France}
\altaffiltext{10}{
Instituto de Astrofisica de Andlucia (CSIC), Camino Bajo de Huetor 50,
18008 Granada, Spain}
\altaffiltext{11}{
Astronomical Institute, Tohoku University, Aoba, Aramaki, Aoba-ku, Sendai
980-8578, Japan}

\begin{abstract}

  We study dust attenuation and stellar mass of $\rm z\sim 0.6$
  star-forming galaxies using new SWIRE observations in IR and GALEX
  observations in UV.  Two samples are selected from the SWIRE and
  GALEX source catalogs in the SWIRE/GALEX field ELAIS-N1-00 ($\Omega
  = 0.8$ deg$^2$). The UV selected sample has 600 galaxies with
  photometric redshift (hereafter photo-z) $0.5 \leq z \leq 0.7$ and
  NUV$\leq 23.5$ (corresponding to $\rm L_{FUV} \geq 10^{9.6}\;
  L_\sun$).  The IR selected sample contains 430 galaxies with
  $f_{24\mu m} \geq 0.2$ mJy ($\rm L_{dust} \geq 10^{10.8}\; L_\sun$)
  in the same photo-z range.  It is found that the mean $\rm
  L_{dust}/L_{FUV}$ ratios of the z=0.6 UV galaxies are consistent
  with that of their z=0 counterparts of the same $\rm L_{FUV}$. For
  IR galaxies, the mean $\rm L_{dust}/L_{FUV}$ ratios of the z=0.6
  LIRGs ($\rm L_{dust} \sim 10^{11}\; L_\sun$)
   are about a factor of 2 lower than local LIRGs, whereas z=0.6
  ULIRGs  ($\rm L_{dust} \sim 10^{12}\; L_\sun$)
  have the same mean $\rm L_{dust}/L_{FUV}$ ratios as their
  local counterparts. This is consistent with the hypothesis that
  the dominant component of LIRG population has changed from large, gas rich
  spirals at z$>0.5$ to major-mergers at z=0. The stellar mass of
  z=0.6 UV galaxies of $\rm L_{FUV} \leq 10^{10.2}\; L_\sun$ is about
  a factor 2 less than their local counterparts of the same
  luminosity, indicating growth of these galaxies.  The mass of z=0.6
  UV lunmous galaxies (UVLGs: $\rm L_{FUV} > 10^{10.2}\; L_\sun$) 
  and IR selected galaxies,
  which are nearly exclusively LIRGs and ULIRGs, is the same as their
  local counterparts.

\end{abstract}


\section{Introduction}
The early results
from rest-frame UV surveys (Lilly et al. 1996; Madau et al. 1996)
reveals an order of magnitude higher star formation rate in $z\sim 1$
galaxies compared to local galaxies. This has been confirmed by recent
large scale UV surveys carried out by Galaxy Evolution Explorer 
(hereafter GALEX, Martin et al. 2005). The
UV luminosity functions for GALEX sources at $z \sim 1$ (Arnouts et al
2005; Schiminovich et al. 2005) and GALEX number counts (Xu et
al. 2005) are consistent with a luminosity evolution index $\alpha
\sim 2.5$ (evolution rate $\propto (1+z)^{\alpha}$).  The IR/sub-mm
surveys reveal a slightly stronger evolution for IR galaxies in the
same redshift range, which is also predominantly luminosity evolution
with an evolution index in the range of $3 \la \alpha \la 4$ (Blain et
al. 1999; Xu 2000; Chary \& Elbaz 2001; Le Floc\'h
2005; Babbedge et al. 2006). 
In an SDSS study of star formation history of local galaxies
(`fossil analysis'), Heavens et al. (2004) concluded that the
peak of star formation of the universe is at z$\sim 0.7$.
Comparing the luminosity density in the UV
(Schiminovich et al. 2005) and that in the FIR (Le Floc\'h et al. 2005)
at different redshifts, the ratio $\rho(FIR)/\rho(FUV)$ increases
by about a factor of 4 from z=0 to z=1, indicating a significant
evolution in the dust attenuation in star forming galaxies during this
epoch (Takeuchi et al. 2005b).

The most significant obstacle preventing accurate measurements
of star formation related quantities is the dust attenuation.
The best way to constrain the dust attenuation is through
the comparison between the UV and infrared emissions (Xu \& Buat
1995; Wang \& Heckman 1996; Meurer et al. 1999; Gordon 2000).
For local galaxies, the UV/IR comparison has been widely
carried out using vacuum UV (100 -- 2000{\AA})
data and IRAS observations (Buat \& Xu 1996;
Wang \& Heckman 1996; Heckman et al. 1998; Iglesias-P\'{a}ramo et
al. 2004). Recent studies using new UV observations obtained
in the GALEX survey (Martin et al. 2005; Buat et al. 2005;
 Iglesias-P\'{a}ramo et al. 2006; Xu et al. 2006) lead to more accurate
estimates for the dust attenuation and its dependence on the
star formation rate, and the selection effects in the UV and IR 
samples. There has been limited analysis of the 
UV/IR comparison of $z>0.4$ galaxies using ISO data and
rest-frame near-UV (2800{\AA}) observations 
(Flores 1998; Hammer et al. 2005), indicating much higher 
dust attenuation in these galaxies than that derived from
UV-optical SED fits (Hammer et al. 2005). Studies of 
Spitzer observations of COMBO-17 galaxies found no evidence
for evolution of the IR-to-UV ratio
versus the star formation rate (SFR) relation over the last
7 Gyrs (Bell et al. 2005; Zheng et al. 2006)

In this paper, we report a study on UV/IR comparisons for 
galaxies of photometric redshifts $0.5 \leq z \leq 0.7$.
Two samples, one UV selected and the other
IR selected, are investigated. 
Galaxies in both samples are star forming galaxies, with
the UV sample including favorably
the galaxies with low dust attenuation and the IR sample
the galaxies with high dust attenuation (Xu et al. 2006;
Buat et al. 2006). We use UV data from the GALEX survey and IR data from the
Spitzer Wide-area Infrared Extragalactic (SWIRE) survey 
(Lonsdale et al. 2004)
to derive the dust attenuation and stellar mass
of these galaxies. Comparisons with 
their local counterparts constrain the evolution of
these properties in the UV and IR selected galaxies, respectively.
In order to take into account the effects due to
different selection functions for the z=0.6 samples
and z=0 control samples,  
the comparisons are  carried out in luminosity bins
where galaxies are found in both the z=0.6 and z=0
samples. 

There is a technical consideration 
for selecting galaxies at $z\sim 0.6$, in addition to the
fact that this redshift is close to the peak of cosmic star formation rate
found by Heavens et al. (2004): At $z\sim 0.6$, the GALEX
NUV band (2350{\AA}) measures the rest-frame FUV band
(1530{\AA}), Spitzer IRAC 3.6$\mu m$ band is close to
rest-frame K band (2.2$\mu m$), and the MIPS 24$\mu m$ band is
close to the rest-frame 15$\mu m$. The rest-frame K band
luminosity is the best indicator of stellar mass (Bell et al.
2003), and the rest-frame 15$\mu m$ luminosity as an
indicator of integrated IR luminosity (5 -- 1000$\mu m$)
has been thoroughly investigated in the context of
ISOCAM 15$\mu m$ observations (Flores et al. 1999;
Franceschini et al. 2001; Chary \& Elbaz et al. 2001).
Therefore comparisons between z=0.6 galaxies and
z=0 galaxies in the corresponding bands will suffer
minimum errors due to the k-corrections which can
be rather uncertain in the far-UV and MIR wavebands.

The paper is organized as following:
After this introduction,
the data sets analyzed in this paper are presented in Section 2.
Major results are listed in Section 3. Systematic uncertainties are
discussed in Section 4. Section 5 and Section 6 are devoted to the
discussion and conclusion, respectively. Through out this
paper, we assume $\Omega_\Lambda = 0.7$, $\Omega_m = 0.3$, and H$_0$ =
70 km sec$^{-1}$ Mpc$^{-1}$.

\begin{figure*}
\plotone{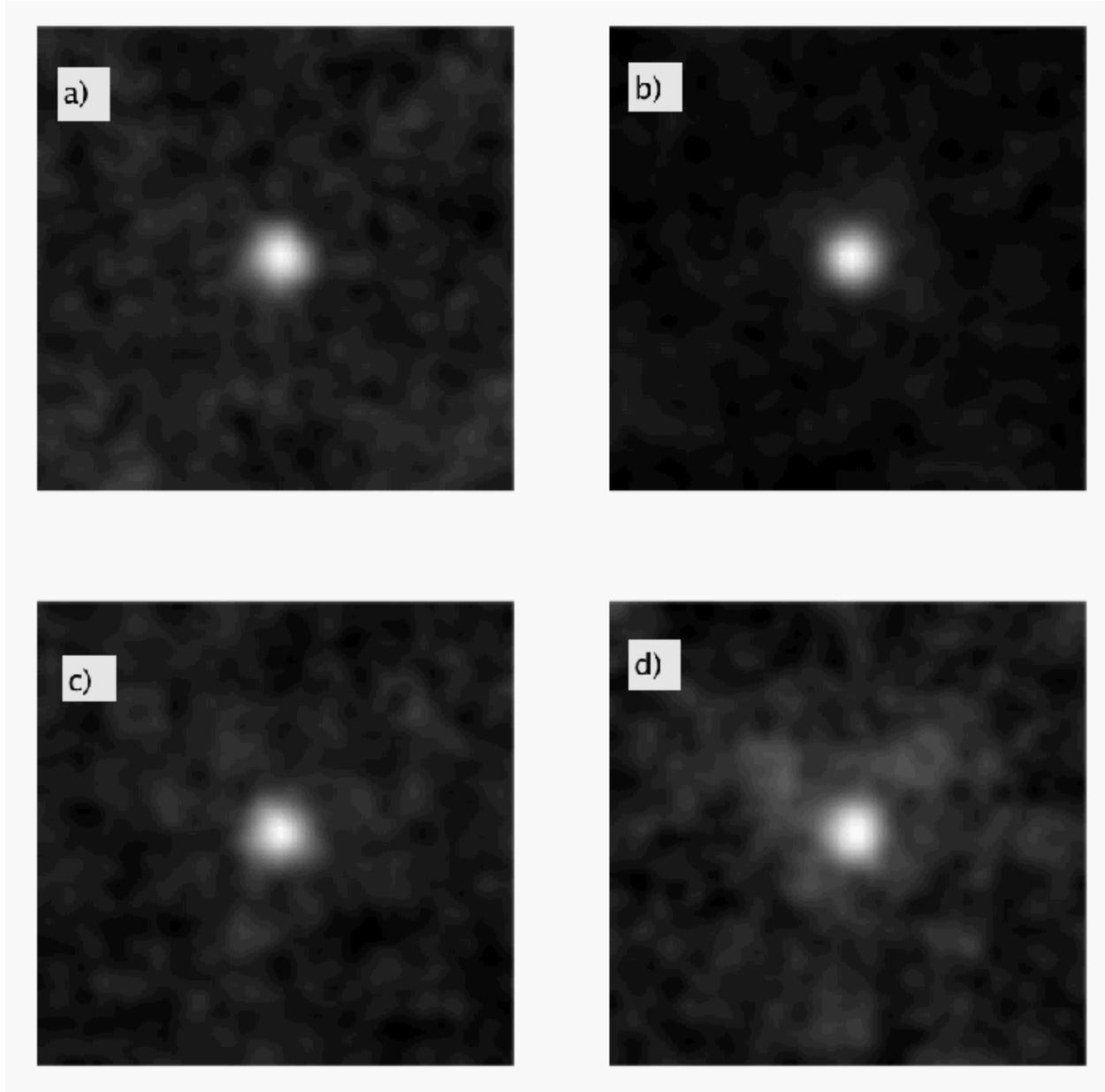}
\caption{Stacked 24$\mu m$ images of z=0.6 UV sources
in the FUV luminosity bins of $9.6 < \log(L_{FUV}/L_\sun) \leq 9.9$ 
({\bf Panel a)}),  $9.9 < \log(L_{FUV}/L_\sun) \leq 10.2$
({\bf Panel b)}), $10.2 < \log(L_{FUV}/L_\sun) \leq 10.5$
({\bf Panel c)}), and $10.5 < \log(L_{FUV}/L_\sun) \leq 10.8$
({\bf Panel d)}).
}
\label{fig1}
\vspace*{-0.2cm}
\end{figure*}

\section{Data and Sample Selection}

\subsection{GALEX and SWIRE data}
The data investigated are in the GALEX field ELAIS-N1-00,
a circular area of 1 deg diameter centered at
RA$=\rm 16^h 13^m 36^s.8$ and Dec=$54^\circ 59' 03".3$, corresponding to
a sky coverage of 0.8 deg$^2$. The field has been observed 
by both GALEX and SWIRE surveys. 
The SWIRE 3.6$\mu m$ and 24$\mu m$ images and fluxes are taken from
``The SWIRE N1 Image Atlases and Source Catalogs'' (Surace et al. 2004).
The nominal 5$\sigma$ sensitivity limit of the SWIRE 24$\mu m$ survey is
$f_{24}=0.2$ mJy, but below $f_{24}=0.25$ mJy the catalog
becomes progressively incomplete (Surace et al. 2004; Shupe
et al. 2006). The GALEX NUV (2350{\AA}) image, a coadd of observations of
8 orbits with a total T$_{exp}$=7899 sec, is taken from
GALEX first public data release (GALEX-DR1). 
NUV sources are extracted from the GALEX image using IRAF DAOPHOT
task. An average foreground extinction of
A$_{NUV}$ = 0.07 mag, estimated using Schlegel map (Schlegel et al. 1998),
 has been corrected. Detailed inspections show
that the 5-$\sigma$ detection reaches NUV=23.5 mag, which is taken
as the flux limit of the NUV catalog. Contaminations
due to false sources become more severe at magnitudes
fainter than this. IR galaxies and UV galaxies undetected either in the 
optical r band (down to r=23.5 mag, Rowan-Robinson et al. 2005)
or in the IRAC 3.6$\micron$ band (with the flux limit 
$f_{3.6\micron} = 3.7 \mu$Jy, Surace et al. 2004)
are excluded in this analysis. 
This significantly reduces the number of false sources in
both bands. Some extreme
populations such as the `extreme 24$\mu m$ galaxies' (Yan et al.
2005; Houck et al. 2005), will be missed
because of the exclusion of sources without optical counterparts.
However, these sources are very rare and most of them are 
hyper-luminous galaxies (HLIRGS)
at high redshift (z$\ga 2$; Houck et al. 2005), outside
the redshift range of our samples. The selection of sources detected
in the IRAC 3.6$\micron$ band should not introduce any
significant bias because observations in this band
is more than an order of magnitude deeper than 
that of the NUV and $f_{24\mu m}$
surveys for an average SED at z=0.6 (see Section 5).
Galaxies with the UV and/or the IR emission dominated by the
active galactic nuclei (AGN) are not explicitly excluded from this study.
It has been shown in the literature that
for both the 24$\mu m$ selected samples and the GALEX selected samples,
the AGN contamination is insignificant, only about 10 -- 15\% of sources
(Franceschini et al. 2005; Bell et al. 2005; Budavari et al. 2005).

\subsection{Photo-z}
The SWIRE photometric redshifts (hereafter photo-z) are derived using
the template-fitting code I{\scriptsize MP}Z (Babbedge et al. 2004;  
Rowan-Robinson 2003).  This method considers a set of galaxy and Type  
1 AGN templates, along with  several priors on dust extinction,  
stellarity and absolute magnitude with redshift in order to obtain  
optimal results.  It has been extended to incorporate IRAC 3.6 and 4.5 
$\micron$ data in addition to optical photometry, 
with the assumption that
the emission in the two IRAC short wavelength bands is dominated
by the stellar radiation (Babbedge et al. 2006; Rowan-Robinson et al. 2005). 
The same ``simple stellar populations'' (SSPs) templates used in
Babbedge et al. (2004), which cover from the UV to the NIR (to $\sim 5\mu m$)
are used. 
The inclusion of the IRAC 3.6 and 4.5 
$\micron$ band data has been shown  
to improve both the reliability and precision of the photo-z's  
(Babbedge et al. 2006; Rowan-Robinson et al. 2005).  In particular,  
the additional NIR data has enabled the rejection of most of the  
extreme outliers resulting from optical-only results and reduced the  
dispersion 
(more details can be found in Section 3.1.2 of 
Babbedge et al. 2006, and in Section 5 of Rowan-Robinson et al. 2005).
Comparisons to spectroscopic redshifts in ELAIS N1 
(P\'{e}rez-Fournon et al. 2006; Serjeant et al. 2006;
Hatziminaoglou et al. 2005) give  
a total $rms$ scatter, $\sigma_{tot}$, of 0.057 for (1+z).
This is consistent with Rowan-Robinson et al (2005) who
quoted an $rms$ of 6.9\% of (1+z) for ELAIS-N1,
where the majority of sources out to  
redshift z$\sim$0.5 are galaxies whilst at higher redshifts (to z$\sim 
$3) Type 1 AGN dominate.  The success of the code has also been  
demonstrated for a number of other fields, filter sets and  
spectroscopic samples (Babbedge et al. 2006).  For all the samples,  
the mean systematic offset between the photometric and spectroscopic  
redshifts was found to be essentially zero to the precision of the  
photometric redshifts.  For example, the ELAIS-N1 sample has $ 
\overline{\Delta} z/(1+z)=0.0037$.

It should be noted that photometric redshift fitting requires 
the photometry in each band being measured in the same way. 
When only using optical data, use
of fixed aperture fluxes is sufficient. However, if we include IRAC
NIR (3.6 and 4.5$\micron$) fluxes it is 
important to be comparing like with like, and
these fluxes are integrated fluxes. Hence, for the optical bands,
the aperture magnitudes corrected to the integrated magnitudes via
curve-of-growth analysis are used. This procedure could introduce
errors for galaxies whose integrated SEDs differ largely from the
SEDs of their central regions, but in practice the results are consistent, 
as found in Babbedge et al. (2006).

Another issue is on the potential mis-identications
between optical sources and IRAC sources due to source confusion, which in
turn might affect the accuracy of the photo-z results.
The accuracy of the cross-IDs of IRAC and optical catalogs of SWIRE sources
has been discussed extensively in Surace et al. (2004). 
The flux limits of both the IRAC and optical catalogs
are well above the confusion limits (Fazio et al. 2004, 
Surace et al., private communication), therefore the probability of
mis-identification due to chance confusion in either band 
is much less than 1\%. Furthermore, the angular resolution of
the optical (FWHM$\sim 1"$ -- $1".5$) and that of the IRAC 3.6 and 4.5 $\micron$
bands (FWHM$\sim 1".6$) are not very different from each other.
Hence, even the genuine close sources (such as galaxy
mergers) tend to have the same status as being resolved or confused 
in both the optical and the IRAC bands in the same time. 
The effect of the confusion on the photo-z results should be insignificant.

\subsection{Samples}
The {\bf UV sample}, selected in the area considered here and in
the photo-z range of $0.5 \leq z \leq 0.7$,
has 600 NUV sources brighter than NUV=23.5 mag.  Among them,
117 are detected by Spitzer at 24$\mu m$ with f$_{24}\geq 0.2$ mJy,
corresponding to an IR detection rate of 20\%. 
For GALEX sources undetected at 24$\mu m$,
upperlimits of $f_{24}=0.2$ mJy are assigned.

The {\bf IR sample} contains
430 SWIRE sources of $f_{24} \geq 0.2$ mJy, selected in
the same area and the same photo-z range ($0.5 \leq z \leq 0.7$). 
Their detection rate by GALEX in NUV is 27\%. 
NUV flux upperlimits
corresponding to NUV=23.5 mag are assigned to
those sources undetected by GALEX.

For both UV and IR samples, 
rest-frame FUV luminosities ($\nu L_\nu (1530{\AA})$) are derived from
the NUV (2350{\AA}) magnitudes and the photo-z.
The Spitzer 24$\mu m$ observations measure the
rest-frame $15\mu m$ emission in these z$\sim 0.6$ galaxies, and the
total dust luminosity is estimated using the conversion
factor $L_{dust}= 11.1\times L_{15}$ (Chary \& Elbaz 2001).

The rest-frame K band (2.2$\mu m$) luminosity
is calculated using $f_{3.6}$
and the photo-z. Stellar mass is estimated
from the K-band luminosity 
using the mass-to-light ratio $\rm M_{stars}/L_K =
0.6 M_\sun/L_\sun$ (Bell et al. 2003), based on a 
Kroupa IMF (Kroupa et al. 1993). This 
is about a factor of 2 lower than the mass-to-light ratio
derived using Salpeter IMF (Cole et al. 2001).

Because we are investigating
galaxies at a given redshift, the
magnitude limit (NUV=23.5 mag) of the UV selected sample 
corresponds to a UV luminosity limit of $L_{FUV}=10^{9.6}$ L$_\sun$
(assuming z=0.6),
and the flux limit of the 24$\mu m$ selected sample
corresponds to an IR luminosity limit of $L_{dust}=10^{10.8}$ L$_\sun$.
Therefore, we are only looking at galaxies in the bright part of
the UV and IR luminosity functions. Particularly, for the IR
selected sample, we study nearly exclusively
LIRGs (L$_{dust} \geq 10^{11}$ L$_\sun$) and 
ULIRGs (L$_{dust} \geq 10^{12}$ L$_\sun$). Given the close to zero
mean systematic offset between the photometric and spectroscopic  
redshifts, our statistical results should be robust
against occasional large errors of the photo-z for some
individual galaxies (`outliers').

The {\bf control samples} at z=0 are taken from Xu et al. (2006),
which are similar to those used in
Buat et al. (2005), Iglesias-P\'{a}ramo et al. (2006).
The UV sample at z=0 includes 94 galaxies brighter than NUV =
16 mag selected from GALEX G1 stage All-sky Imaging Survey (AIS),
covering 654 deg$^2$. The z=0 FIR sample includes 161  galaxies with $f_{60} \geq
0.6$ Jy in 509 deg$^2$ sky covered both by GALEX AIS and IRAS PSCz
(Saunders et al. 2000). More details can be found
in Xu et al. (2006), Buat et al (2005) and Iglesias-P\'{a}tramo et al. (2006).
Since many quantities (such as the
dust attenuation) have strong luminosity dependence,
we always compare the z=0.6 and
z=0 galaxies in the same luminosity bins.

\section{Results}

\subsection{Dust Attenuation in z=0.6 Galaxies}  
\subsubsection{UV Galaxies at z=0.6}  
The 24$\mu m$ detection rate of the z$= 0.6$ NUV sources
is only 20\% (Section 2). Therefore for a large majority
of these sources the MIR emission is below the SWIRE
sensitivity limit. In order to derive meaningful statistics
related to the infrared emission, we carried out stacking analysis
for galaxies binned into four UV luminosity bins (Table 1).

For each NUV galaxy included in a bin, we cut from the background-subtracted
24 $\mu$m mosaic a small subimage, 1 arcminute on a side, and centered
on the coordinates given in the GALEX catalog.  From a stack of these
subimages, we compute a trimmed mean image, excluding 20\% of the subimages
with the lowest and highest brightness (10\% at either end).  The
purpose of the trimming is to guard against contamination from nearby
bright sources, as well as mis-classified sources with large
photo-z errors (`outliers').  
The mean flux density is measured from the trimmed mean
image in an aperture of 18\arcsec diameter.  To estimate the
standard error of the mean flux density, we use a bootstrap method.
From the stack of subimages, we sample with replacement the
subimages and make a new trimmed mean image, and measure the flux
density in the aperture.  The resampling is repeated 1000 times,
and the sample standard deviation of the aperture measurements
provides the uncertainty estimate.
The average $\log \rm (L_{dust}/L_{FUV})$ is derived from the mean
$f_{24}$, mean redshift z=0.6, and the mean $L_{dust}$ of the bin.
The error of mean $\log \rm (L_{dust}/L_{FUV})$ is the quadratic sum of
the error of $f_{NUV}$ and that of $<L_{dust}>$.
The stacked (trimmed mean) images are shown in
Fig.1. The results are reported in Table 1.

\begin{figure*}
\plotone{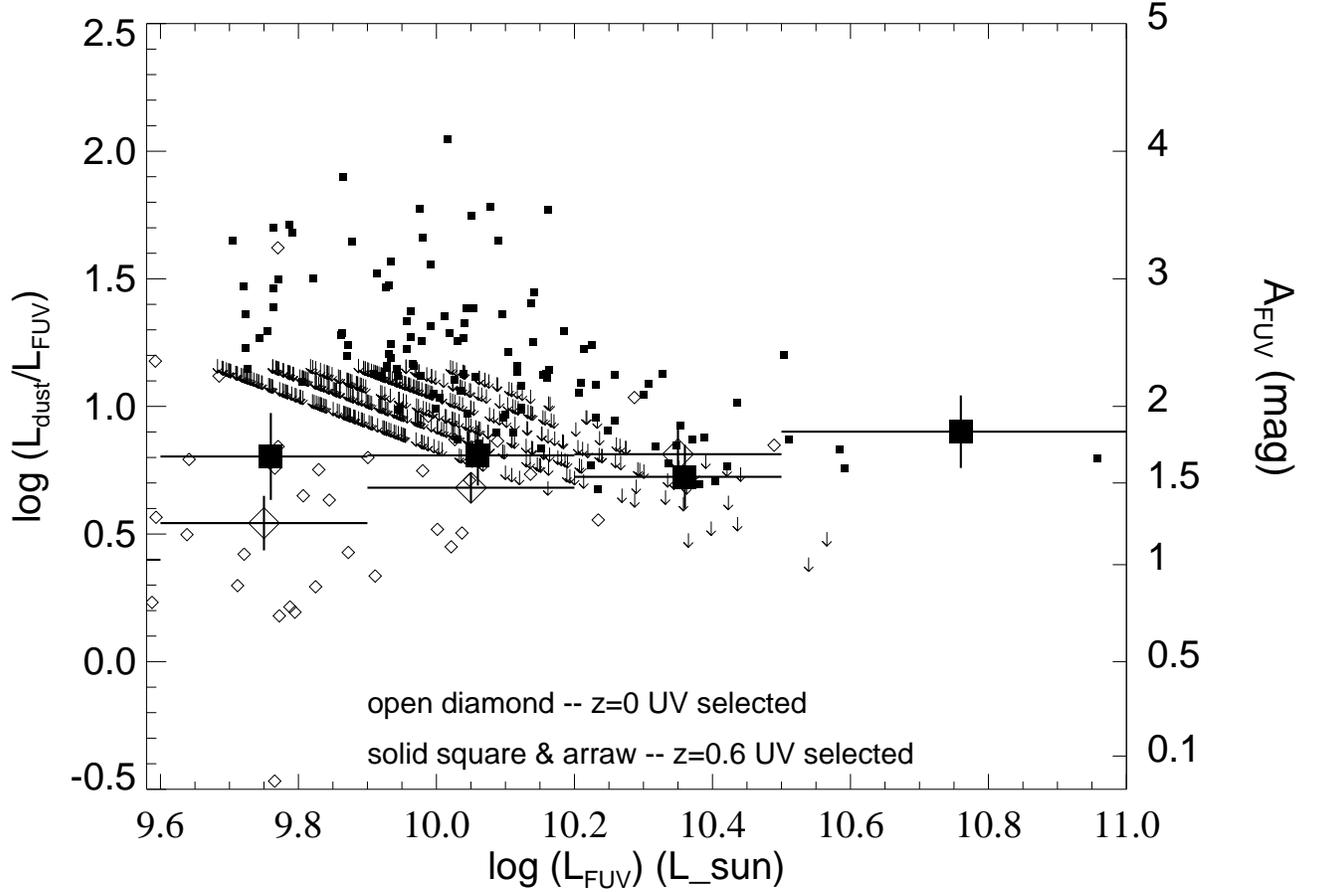}
\caption{log (L$_{dust}$/L$_{FUV}$) v.s. log(L$_{FUV}$) plot for the 
UV selected samples at z=0.6 and z=0. Arrows denote upperlimits.
Large solid squares with error bars are the means for z=0.6
galaxies, derived through stacking analysis.
Large open diamonds with error bars are the means for z=0
galaxies. The right hand axis of the plot marks the FUV attenuation (A$_{FUV}$)
corresponding to the log(L$_{dust}$/L$_{FUV}$).
} 
\end{figure*}

\begin{deluxetable}{ccccccc}
\tablewidth{0pt}
\tablecaption{Mean IR fluxes and IR-to-UV ratios of
z=0.6 UV galaxies.}
\tablehead{
\colhead{log L$_{FUV}$} & \colhead{N$_{tot}$} 
& \colhead{N$_{det}$}  & \colhead{$f_{24}^{stack}$} 
& \colhead{$\sigma$} & \colhead{log(L$_{dust}$/L$_{FUV}$)} 
& \colhead{error} \\ 
\colhead{(L$_\sun$)} & & & \colhead{($\mu$Jy)} & \colhead{($\mu$Jy)} & & }
\startdata
 9.75$\pm$0.15   & 187  & 21 & 67  & 25 &  0.80 & 0.17\\
10.05$\pm$0.15   & 341 & 68 & 135 & 21 &  0.80 & 0.12 \\
10.35$\pm$0.15 & 65  & 24 & 222 & 42 &  0.72 & 0.13 \\
10.65$\pm$0.15 & 7   & 5  & 838 & 219 & 0.90 & 0.14 \\

\enddata
\end{deluxetable}

%


In Fig.2 we compare the L$_{dust}$/L$_{FUV}$ ratios
of the z=0.6 UV galaxies with
those of the z=0 UV galaxies in the control sample. 
The FUV attenuation (A$_{FUV}$) values corresponding to given 
log(L$_{dust}$/L$_{FUV}$), marked on the right hand axis of the plot,
are calculated using the following formula taken from Buat et al. (2005):
\begin{equation}
\rm A_{FUV}=-0.00333 y^3 + 0.3522 y^2 +1.1960 y + 0.4967,
\end{equation}
where $\rm y=\log (L_{dust}/L_{FUV})$.
Small solid squares and arrows represent individual z=0.6 galaxies, 
and small open diamonds the z=0 galaxies. The arrows, denoting the upper-limits
in the z=0.6 sample, concentrate in a narrow, tilted region. This is because
the redshifts (photo-z) of the galaxies in this sample are in a very narrow
range of $0.5 \leq z \leq 0.7$. Therefore the 25$\mu m$ upper limits
(0.2 mJy) are translated to a narrow range of $\rm L_{dust}$ upper limits.
The large symbols with error bars are the mean ratios in the corresponding
luminosity bins. For the z=0.6 sample, they are calculated from the average
25$\mu m$ fluxes measured on the stacked images. As demonstrated in Fig.1,
good detections are obtained in all four stacked 24$\mu m$ images. Therefore,
unlike for ratios of individual galaxies, the mean  L$_{dust}$/L$_{FUV}$ ratios
are not affected by sensitivity limit of the SWIRE survey.
 For the z=0 sample, since all individual 
sources are detected in the IR band, the calculations of the means
are straightforward.

For both z=0.6 and z=0 samples, the  L$_{dust}$/L$_{FUV}$ ratio
does not show significant dependence on the UV luminosity.
It is interesting to note that many z=0.6 UV galaxies
with $\rm \log(L_{FUV}) < 10.2$
have $\rm \log(L_{dust}$/L$_{FUV}) > 1.5$, corresponding
to the FUV attenuation $\rm A_{FUV} \ga 3$ mag (Buat et al. 2005), whereas
such high dust attenuation is seldomly seen in their
local counterparts. In the UV luminosity bin of $9.6 \leq \log (L_{FUV}/L_\sun)
< 9.9$, the mean L$_{dust}$/L$_{FUV}$ ratios of z=0 and z=0.6 galaxies
show a $\sim 80$\% (i.e. 0.25 dex) difference at $\sim 1\sigma$ level.
In the remaining two luminosity bins where the
two samples overlap,
the mean L$_{dust}$/L$_{FUV}$ ratios of z=0 and z=0.6 galaxies are very close
to each other (difference $< 50\%$).
Recently, Burgarella et al. (2006) found evidence for
about half of
z$\sim 1$ UV bright galaxies to have very low 
dust attenuation ($\rm A_{FUV} \sim 0.5$ -- 0.6 mag).
We did not detect the same trend for  $z\sim 0.6$ UV galaxies.

\begin{deluxetable}{ccccccc}
\tablewidth{0pt}
\tablecaption{Mean UV fluxes and IR-to-UV ratios of z=0.6 IR galaxies.}
\tablehead{
\colhead{log(L$_{dust}$)} & \colhead{N$_{tot}$} 
& \colhead{N$_{det}$}  & \colhead{$NUV^{stack}$}  & \colhead{$\sigma$} 
& \colhead{log(L$_{dust}$/L$_{FUV}$)}
& \colhead{error} \\ 
\colhead{(L$_\sun$)} & & & \colhead{(mag)} & \colhead{(mag)} & & }
\startdata
11.0$\pm$0.2   & 172  & 52 & 24.15 & 0.20& 1.37 & 0.11 \\
11.35$\pm$0.15 & 200 & 50 & 24.30 & 0.18 & 1.74 & 0.11 \\
11.65$\pm$0.15 & 49  & 12 & 24.22 & 0.28 & 2.04 & 0.12 \\
12.0$\pm$0.2   & 9  & 3   & 23.76 & 0.39 & 2.25 & 0.16 \\
\enddata
\end{deluxetable}


\subsubsection{IR Galaxies at z=0.6}  

For the 430 z=0.6 galaxies in the 24$\micron$ selected
sample, the detection rate in the GALEX NUV band is also low (27\%). 
Therefore the stacking analysis is again exploited in deriving
the mean NUV fluxes. Galaxies are binned into four IR luminosity
bins (Table 2). Because the depth of the NUV image is very
close to the confusion limit (NUV=24 mag, Xu et al. 2005),
we choose to subtract sources brighter than NUV=23.5 mag
(5$\sigma$ detections) from the image before stacking. 
In principle this approach
should yield a cleaner result (Zheng et al. 2006)
than simply stacking all sources
because the contamination due to bright neighboring sources outside
the sample (i.e. sources of different redshifts) is minimized.
Again, we derive the trimmed mean
of the NUV flux for each bin of IR selected galaxies 
by excluding 20\% of galaxies with the highest and the lowest measured
NUV fluxes (10\% on each side), and estimate the errors
by bootstrapping (1000 replicate simulations). For each  $L_{dust}$ bin,
the average $\log \rm L_{dust}/L_{FUV}$ and its error are derived
in the same way as for UV galaxies in a given $L_{FUV}$ bin.
The results are reported in Table 2. 

\begin{figure*}
\plotone{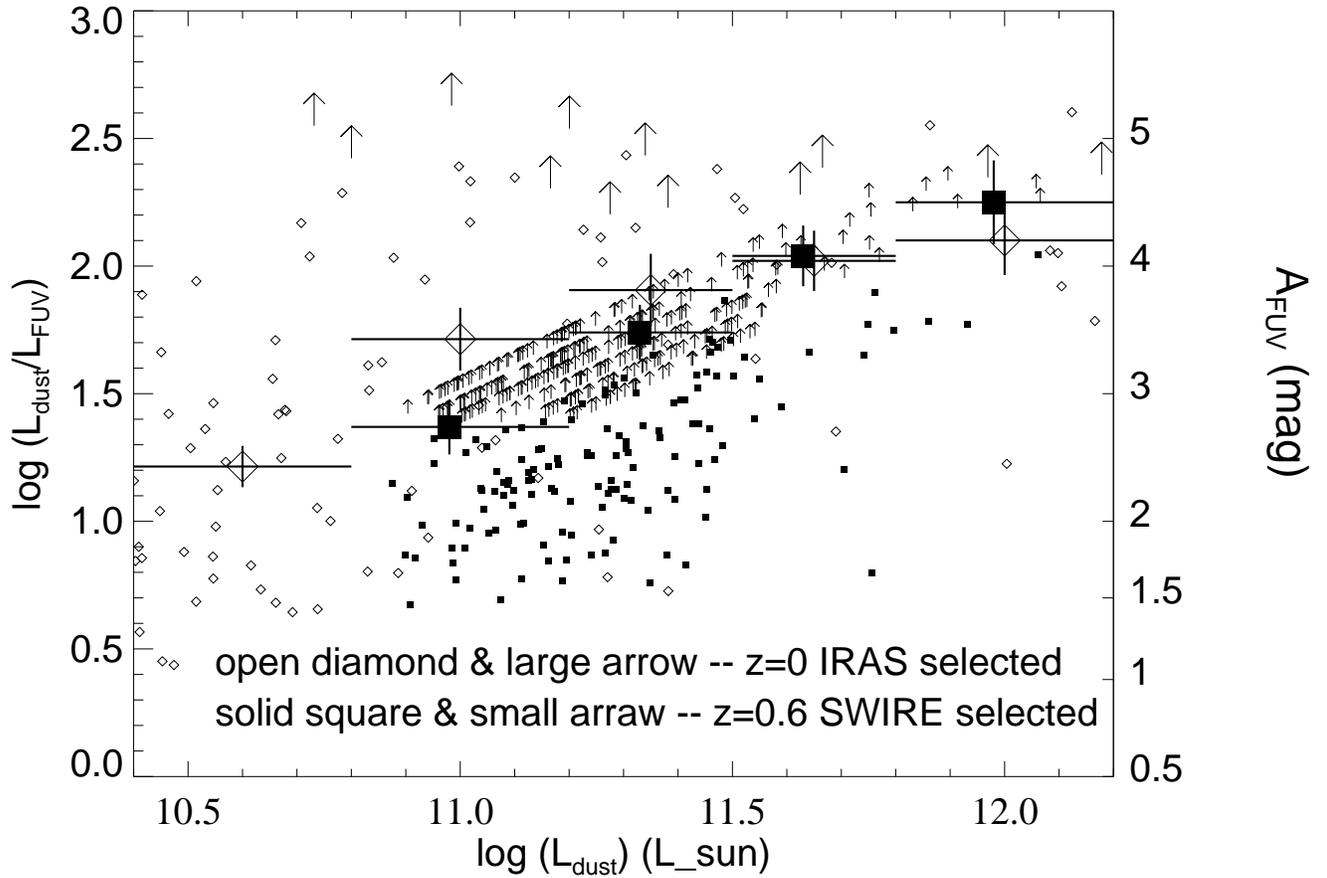}
\caption{L$_{dust}$/L$_{FUV}$ v.s. L$_{dust}$ plot for the 
IR selected samples at z=0.6 and z=0. Arrows denote upperlimits.
The large solid squares with error bars are the means for z=0.6
galaxies.
The large open diamonds with error bars are the means for z=0
galaxies.
} 
\end{figure*}

In Fig.3 we compare the L$_{dust}$/L$_{FUV}$ ratios of the z=0.6 IR
galaxies with those of their z=0 counterparts.
Same as in Fig.2, small symbols and arrows represent individual galaxies,
and large symbols with error bars the means in corresponding luminosity
bins. In contrast with the UV galaxies in Fig.2,
both z=0.6 and z=0 IR galaxies show strong dependence of the
L$_{dust}$/L$_{FUV}$ ratio with the luminosity. 
In the luminosity range covered by the $z=0.6$ sample ($\ga 10^{11} L_\sun$),
the L$_{dust}$/L$_{FUV}$ ratios of IR galaxies are one to two orders of
magnitude higher than those of the UV galaxies
(Fig.2).  It appears that
z=0.6 IR galaxies have a steeper slope in the 
$\rm log (L_{dust}/L_{FUV})$ versus $\rm log (L_{dust})$ relation than
z=0 IR galaxies.  The mean
L$_{dust}$/L$_{FUV}$ of z=0.6 galaxies in the bin $\rm 10.8\leq
log(L_{dust}/L_\sun) <11.2$ is significantly lower, by a factor of $\sim
2.5$, than that of their local counterparts, indicating 
that z=0.6 LIRGs ($\rm log (L_{dust}/L_\sun)
\sim 11$) have lower dust attenuation compared to their local
counterparts. On the other hand, as shown in Fig.3,
z=0.6 ULIRGs ($\rm log (L_{dust}/L_\sun) \sim 12$)
have similar dust attenuation as local ULIRGs.

Bell et al. (2005) and Zheng et al. (2006),
in their studies of MIPS observations of
COMBO-17 galaxies, concluded that there is no evolution in
the L$_{dust}$/L$_{FUV}$ versus SFR relation over the last 7 Gyr.
It is not straightforward to compare their results with ours because
our sample is IR selected whereas COMBO-17 is an
optical sample. The analysis of Bell et al. (2005)
is based on a comparison of the L$_{dust}$/L$_{FUV}$ versus SFR
plot of individual z$\sim 0.7$ galaxies with that of their local counterparts
(Fig.1 of Bell et al. 2005),
which can be affected by the presence of large fraction ($\sim 70\%$) 
of upperlimits in the MIPS data. In Fig.9 of Zheng et al. (2006), 
only galaxies in the top two panels (corresponding to the $\rm M_B$ bins
of $\rm [M_*-1, M_*]$ and $\rm [M_*, M_*+1]$) include LIRGs
(SFR $\rm > 10  M_\sun$ yr$^{-1}$). For these galaxies, there is indeed a trend
that while the average SFR increases by about an order of magnitude from
z=0.15 to z=0.95, the L$_{dust}$/L$_{FUV}$ ratio increases very little,
at least significantly less than what is predicted by 
the  L$_{dust}$/L$_{FUV}$ versus SFR relation.
Le Floc'h et al (2005) studied a MIPS selected sample (median redshift
$\sim 0.7$) among COMBO-17 galaxies. Inspections of their 
$\rm L_{IR}/L_{UV}$ versus $\rm L{IR}$ plot (their Fig.10c) show
that the median $\rm L_{IR}/L_{UV}$ of the LIRGs in that sample
is also significantly lower than that of local LIRGs, consistent with
our result.

\subsection{Stellar Mass of z=0.6 Galaxies}

\subsubsection{UV Galaxies at z=0.6}  
\begin{figure*}
\plotone{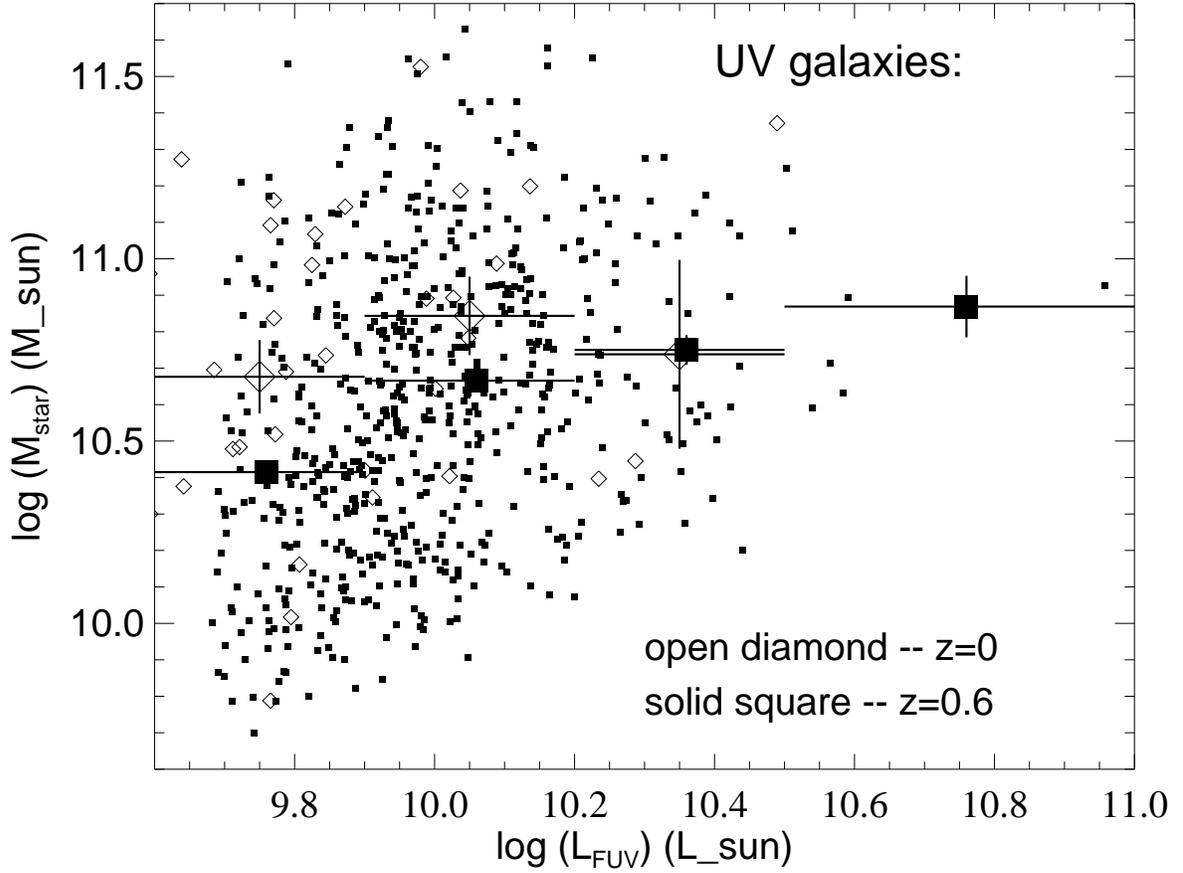}
\caption{Plot of stellar mass (estimated using rest-frame K band
luminosity) vs. FUV luminosity for UV selected samples at 
z=0.6 (solid squares) and z=0
(open diamonds). The small symbols are individual galaxies,
whereas the large symbols with error bars are the corresponding means.
} 
\end{figure*}

The stellar mass of z=0.6 galaxies is estimated using the
IRAC 3.6$\mu m$ flux, which measures the rest-frame
K-band emission (Section 2.2). 
For the control samples at z=0,
the stellar mass is estimated using the total magnitude ($\rm m_{tot}$)
of the 2MASS K$_s$ (2.16$\mu m$) band (Jarrett et al. 2000).
Interestingly, as shown in Fig.4, 
the stellar mass of z=0.6 UV galaxies fainter than 
$\rm L_{FUV}=10^{10.2}$ L$_\sun$ is on average a factor
of 1.5 to 2 lower than that of their local counterparts.
This result is statistically significant at the 2$\sigma$
level. These galaxies are in the category of `intermediate
massive galaxies' ($3\times 10^{10}$ M$_\sun$ $\leq M \leq 3\times 10^{11}$
M$_\sun$) as defined by Hammer et al. (2005). According to Heavens
et al. (2004), the star formation in these galaxies peaked
at $z\sim 0.6$. Hammer et al. (2005) argued that the mass of
these galaxies increased by about a factor of 2 since z=1.
Our result is consistent with this.
On the other hand, the z=0.6 UV luminous galaxies
(UVLGs, Heckman et al. 2005) of 
$\rm L_{FUV} > 10^{10.2}$ L$_\sun$ have the same mean
stellar mass as their local counterparts.

\begin{figure*}
\plotone{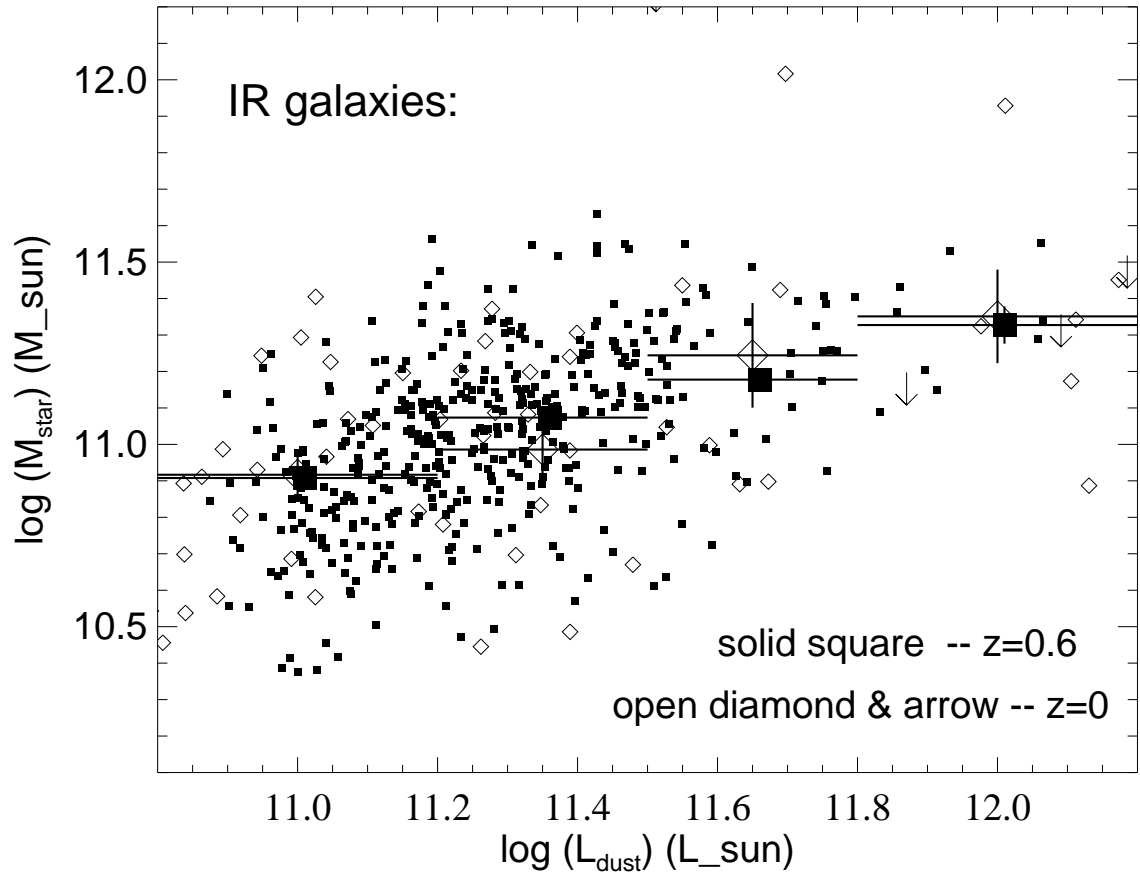}
\caption{Plot of stellar mass vs. $L_{dust}$ for IR selected samples at z=0.6
(solid squares) and z=0 (open diamonds).  
The small symbols are individual galaxies,
whereas the large symbols with error bars are the corresponding means.
} 
\end{figure*}

\subsubsection{IR Galaxies at z=0.6}  
In Fig.5 we compare the stellar mass of z=0.6 IR galaxies 
with that of their local counterparts.
Here, unlike for UV galaxies, no systematic
difference is found between the means of z=0.6 and
z=0 galaxies. Both samples show the same trend that
more luminous galaxies have higher stellar mass.

It should be pointed out that for
bright IR galaxies, particularly the ULIRGs,
the contribution from the violent starburst 
to the rest-frame NIR emission can
be significant (Surace et al. 2000), accounting for up to
$\sim 50\%$ of the K band flux. Therefore the
stellar mass estimated using the rest-frame K-band
should be treated with caution. On the other hand,
the conclusion derived from Fig.5 should
not be affected by this if the contamination from the starburst is
the same for the z=0 and for the z=0.6 galaxies
of the same luminosity.


\subsubsection{Comparison between UV and IR Galaxies}
UV and IR samples select low and high dust attenuation galaxies, 
respectively (Xu et al. 2006; Buat et al. 2006). Given the
dependence of the dust attenuation on stellar mass
(Wang \& Heckman 1996; Burgarella et al. 2005), the UV galaxies
tend to have lower stellar mass than the IR galaxies. This is indeed
what we see in Fig.4 and Fig.5. In particular, no IR galaxy in
the z=0.6 sample has the stellar mass less than $10^{10.3}$ M$_{\sun}$,
while as many as about 20\% of 
galaxies in the z=0.6 UV sample have stellar mass below this.
It should be noted that the IRAC 3.6$\micron$ band 
flux limit $f_{3.6\micron} = 3.7\mu$Jy corresponds to a stellar mass limit
of $M = 10^{9.4}$ M$_\sun$, a factor of $\sim 2$ ($\sim 8$) lower
than the minimum stellar mass of the z=0.6 UV galaxies (IR galaxies).
This demonstrates again that very few (if any) galaxies might have
been missed by our UV and IR samples due to the IRAC 3.6$\micron$ 
flux limit.


UVLGs in the last two UV luminosity bins in Fig.4 have their mean $\rm
L_{dust} > 10^{11}$ L$_\sun$ (Table 1), therefore belonging to the
LIRG population as well.  These are galaxies with the most active star
formation in the universe, likely being in the peak phase of some
brief (and perhaps recurrent) starburst episodes (Hammer et al. 2005).
In the literature, three mechanisms have been considered for the
triggering of starbursts, including (1) major-merger, (2) minor
merger, (3) bar instability during the secular evolution of disk
galaxies (Hammer et al. 2005; Bell et al. 2005; Combes 2006). There is
indication that relative importance of these mechanisms has changed
since $z \sim 1$ (Melbourne et al. 2005).  Our results in Fig.4 and
Fig.5 indicate that, no matter which mechanism dominates the
triggering of LIRG/ULIRG activity, the host galaxies of these
starburst events in the z=0.6 and z=0 universe have the same stellar
mass.


\section{Systematic Uncertainties}

\subsection{The  L$_{dust}$/L$_{15}$ ratio}
\begin{figure*}
\plotone{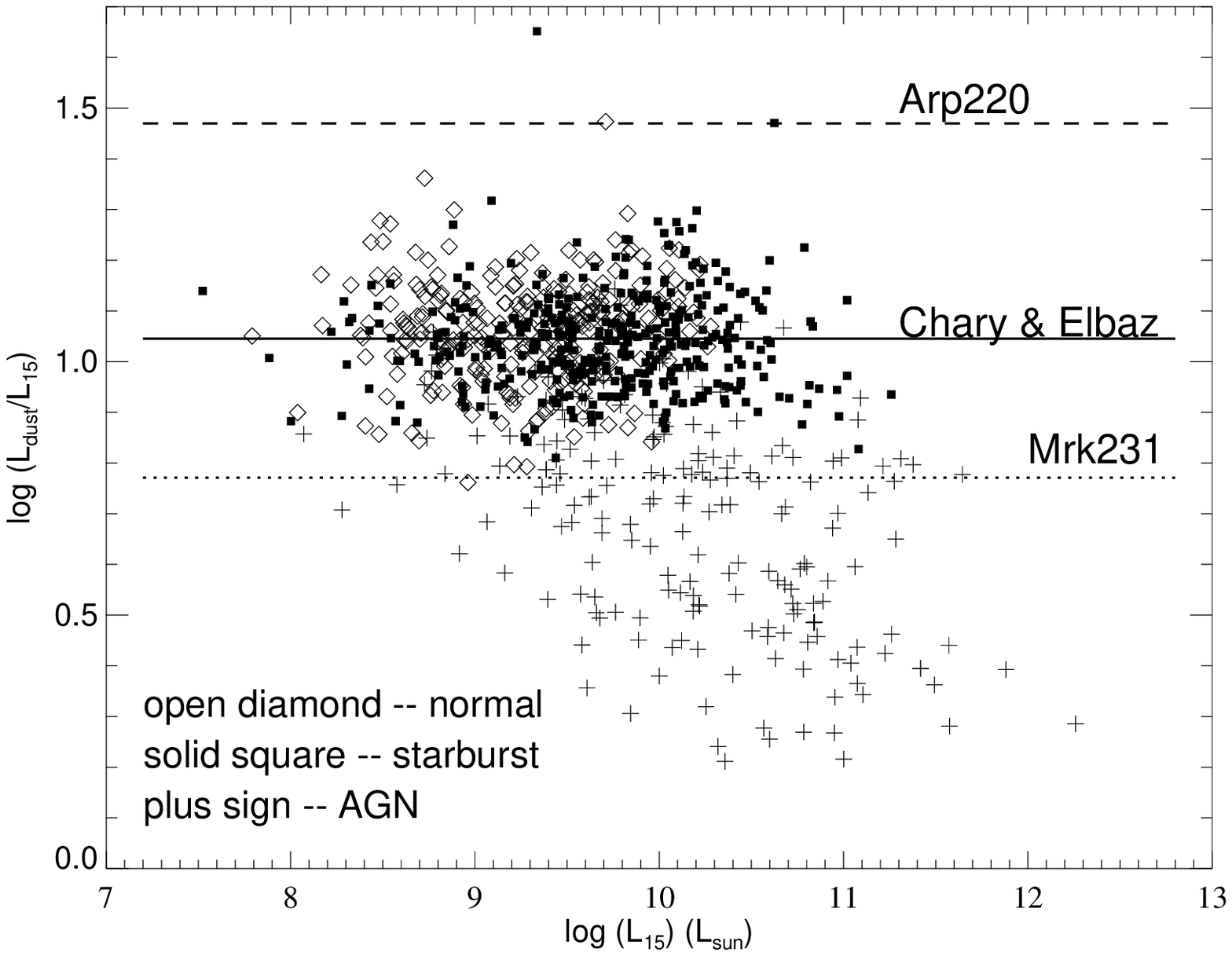}
\caption{L$_{dust}$/L$_{15}$ v.s. L$_{15}$ plot for 
local IR galaxies in the SED library of Xu et al. (2001).}
\end{figure*}

\begin{figure*}
\plotone{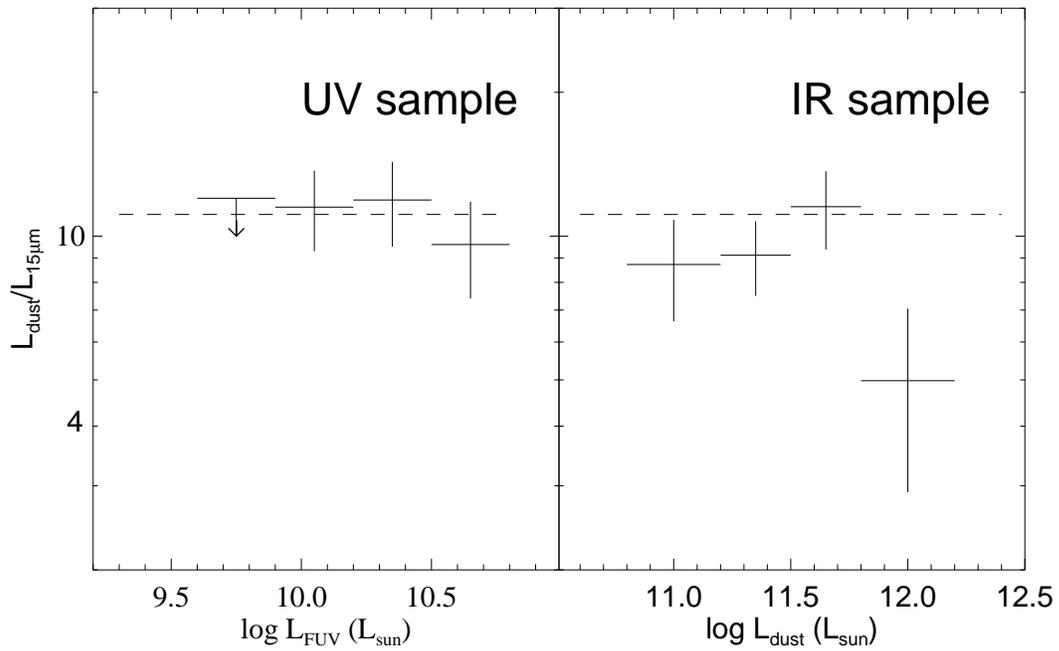}
\caption{Mean L$_{dust}$/L$_{15}$ ratios for z=0.6 UV galaxies
(left panel) and for z=0.6 IR galaxies (right panel), estimated
using the mean $f_{160}/f_{24}$ and $f_{70}/f_{24}$ ratios derived
by stacking. The dotted line specifies the adopted standard taken
from Chary \& Elbaz (2001).} 
\end{figure*}

\begin{figure*}
\plotone{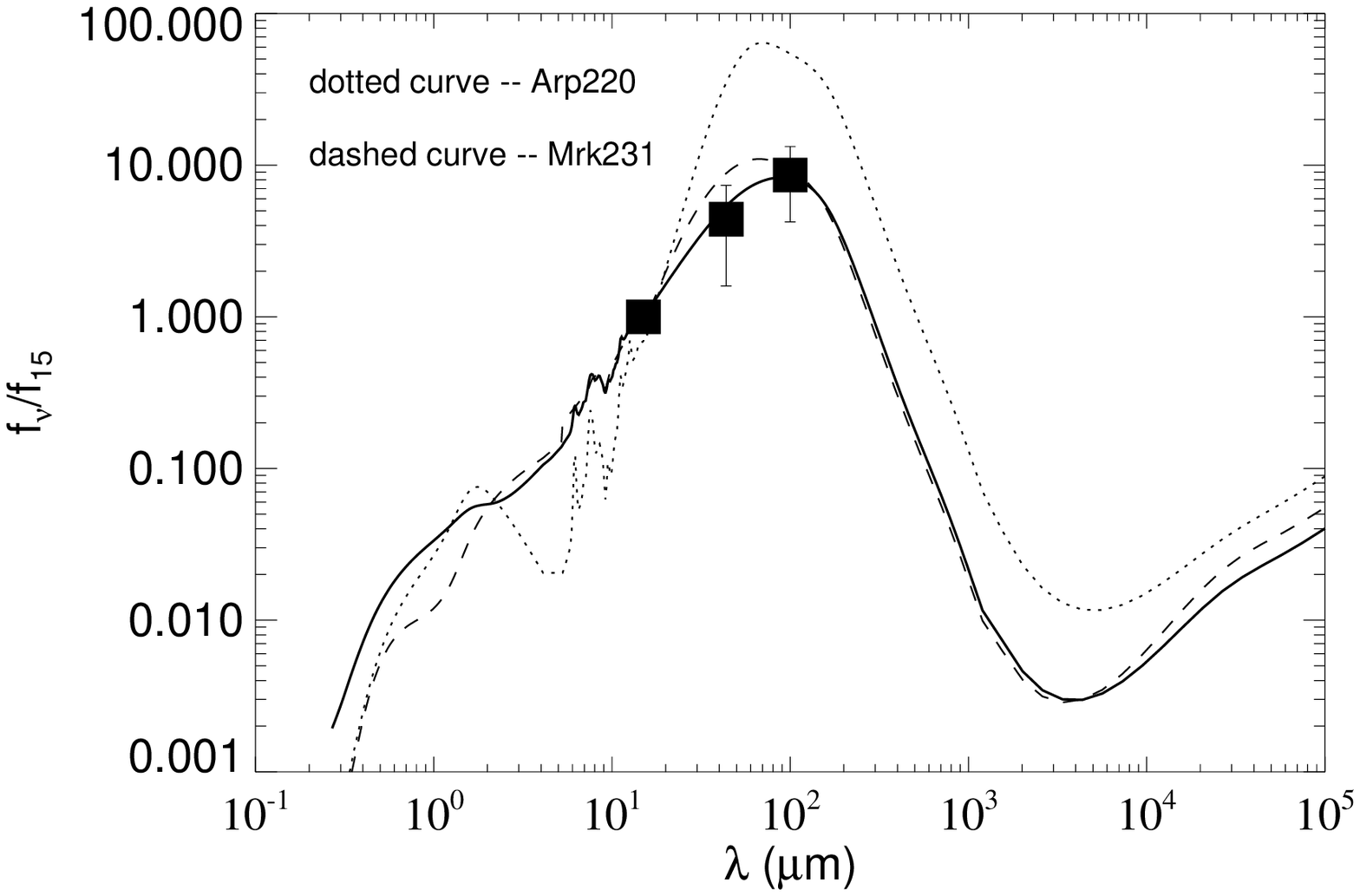}
\caption{Plot of the rest-frame SED (solid curve) that fits the mean
FIR color ratios (solid squares with error bars) of z=0.6 ULIRGs (IR
galaxies of $\rm 11.8\leq \log (L_{dust}/L_\sun) <12.2$). Compared
to the SED of Arp~220 (dotted curve) and that of Mrk~231 (dashed
curve).}
\end{figure*}

The most important source of systematic uncertainty in this work is
due to the extrapolation from $\rm L_{15}$ to L$_{dust}$.  The
analysis of SEDs local IR galaxies by Chary \& Elbaz (2001) has shown
that L$_{dust} \propto$ L$_{15}^{0.998\pm 0.021}$, therefore the
L$_{dust}$/L$_{15}$ ratio has little dependence on the luminosity (see
also Takeuchi et al. 2005a). We checked this result using the IR SEDs
of a larger sample of 831 IRAS galaxies that are constructed
`semi-empirically' by Xu et al. (2001).  The result is plotted
in Fig.6.  Xu et al. (2001) divided the SED sample into 3 sub-classes
according to the IRAS colors: normal disk galaxies ($\rm f_{60}/f_{25} > 5$
and $\rm f_{100}/f_{60} \geq 2$), starburst
galaxies  ($\rm f_{60}/f_{25} > 5$
and $\rm f_{100}/f_{60} < 2$), and AGNs  ($\rm f_{60}/f_{25} \leq 5$). 
As shown in Fig.6, indeed for normal disks and
starbursts the L$_{dust}$/L$_{15}$ ratio is rather constant against
the luminosity. The mean L$_{dust}$/L$_{15}$ of normal disk galaxies
is 11.5$\pm 3.1$ and that of starburst galaxies is 11.0$\pm 2.8$, both very
close to the value of Chary \& Elbaz (2001) which is 11.1. On the other
hand, there is a significant trend for galaxies with AGN in the sense
that the ratio decreases with L$_{15}$.  The mean of IR galaxies with
AGN is 4.9$\pm 3.8$, about a factor 2 lower than those of normal disk
and starburst galaxies. Therefore, we might have significantly over-estimated
the L$_{dust}$ if galaxies in any of the luminosity bins studied here are
dominated by AGNs. In Fig.6, the ratios of two famous
ULIRG Arp~220 (29.5) and Mrk~231 (5.9), the former a prototype of `cold
ULIRGs' which are mostly starbursts and the latter a prototype of
`warm ULIRGs' mostly AGNs, are also plotted as references.
In addition to the intrinsic variations in the L$_{dust}$/L$_{15}$
ratio, the effect of possible SED evolution in z=0.6 galaxies has to
be taken into account, too. 

The most direct way to constrain this
uncertainty is to look at the real SEDs of the z=0.6 galaxies
detected in longer wavelength bands of Spitzer, in particular 
the MIPS 160$\mu m$ (rest-frame 100$\mu m$) band. However,
it turned out that only 1 source in the z=0.6 IR sample
is detected at 160$\mu m$ above the 
nominal 5$\sigma$ sensitivity limit of $f_{160}=100$ mJy
(Surace et al. 2004). A much more robust method is to
stack the 70$\mu m$ and 160$\mu m$ images of the galaxies
in a given luminosity bin and use the mean
$f_{70}/f_{24}$ and $f_{160}/f_{24}$
derived from the stacked images
to constrain the mean SED. The 70$\mu m$ and 160$\mu m$ images
are taken from the same SWIRE database.  For each luminosity
bin, SEDs taken from
the sample plotted in Fig.6 are selected according to the
IR luminosity range, the mean $f_{70}/f_{24}$ and $f_{160}/f_{24}$
and their uncertainties (including 20\% calibration uncertainty).
The mean
L$_{dust}$/L$_{15}$ ratio and the uncertainty are derived using these SEDs.
The results are given in Table~3 and Table~4, and plotted in Fig.7.
Note that for UV galaxies in the luminosity bin of $9.6 \leq \log (L_{FUV}/L_\sun)
 < 9.9$, no detections are found even on the stacked 70$\mu m$ and
160$\mu m$ images, therefore only upperlimits are listed.
For galaxies in other luminosity bins in both Table~3 and Table~4,
the detections on stacked 70$\mu m$ and 160$\mu m$ images 
are significant ($>3$ times of the noise), although some of the errors
of the mean flux ratios (particularly those of
$\rm f_{160}/f_{24}$) are as high as 90\%. In these cases the
error of the mean is largely due to the statistical dispersion of the 
variable.

For all UV selected galaxies, 
the derived mean
L$_{dust}$/L$_{15}$ ratios for all the 4 luminosity bins
are consistent with that of Chary and Elbaz (2001).
For IR selected galaxies, only the mean ratio of galaxies 
in the brightest bin of $\rm 11.8\leq \log L_{dust} <12.2$ 
is significantly below the calibration of Chary and Elbaz (2001).
As shown in Fig.8, the SED which fits the mean FIR colors of the 
bin is much closer to that of Mrk~231 than Arp~220, suggesting that 
many galaxies in this bin have their 24$\mu m$ flux enhanced 
by AGN dust torus emission.
This can be compared with the ISO results
(Genzel et al. 1998) which show that most of the IRAS 60$\mu m$ band selected
z=0 ULIRGs of $\rm  L_{dust} \sim 10^{12} L_\sun$
are powered by starbursts, and AGN contribution is important only
for ULIRGs with $ \rm L_{dust} > 2\; 10^{12}  L_\sun$ (Veilleux et al. 1998).
The increased contribution in the MIR of AGN in the ULIRGs
of our z=0.6 IR sample
is at least partly a consequence of the MIR (rest-frame 15$\mu m$)
selection. Whether this also indicates a difference between
z=0 and z=0.6 ULIRGs requires further investigation.
Meanwhile, as shown in Fig.3, if the mean $\rm L_{dust}$ of the z=0.6 galaxies
in the last luminosity bin is reduced by a factor of 2, their 
$\rm L_{dust}$/L$_{FUV}$ ratio is still consistent with that of the local
ULIRGs.

\begin{deluxetable}{ccccccc}
\tablewidth{0pt}
\tablecaption{Mean FIR colors and L$_{dust}$/L$_{15}$ ratios
of z=0.6 UV galaxies.}
\tablehead{
\colhead{L$_{FUV}$} 
& \colhead{$f_{70}/f_{24}$}  & \colhead{$error$} 
& \colhead{$f_{160}/f_{24}$}  & \colhead{$error$} 
& \colhead{$\rm L_{dust}/L_{15}$}  & \colhead{error} \\ 
\colhead{(L$_\sun$)} & & & & &&}
\startdata
 9.75$\pm$0.15   & $<11.4$ & ... & $<24.4$ & ... & $< 12.0$ & ... \\ 
10.05$\pm$0.15   &  6.1  & 3.0 & 27.8 & 20.2 & 11.5 & 2.2 \\
10.35$\pm$0.15 & 10.3  & 4.7 & 38.9 & 25.0  & 11.9 & 2.4 \\
10.65$\pm$0.15 & 4.1   & 3.0 & 19.3 & 14.2  & 9.6 & 2.2 \\
\enddata
\end{deluxetable}

\begin{deluxetable}{ccccccc}
\tablewidth{0pt}
\tablecaption{Mean FIR colors and L$_{dust}$/L$_{15}$ ratios
of z=0.6 IR galaxies.}
\tablehead{
\colhead{L$_{FUV}$}  
& \colhead{$f_{70}/f_{24}$}  & \colhead{$error$} 
& \colhead{$f_{160}/f_{24}$}  & \colhead{$error$} 
& \colhead{$\rm L_{dust}/L_{15}$}  
& \colhead{error} \\ 
\colhead{(L$_\sun$)}& & & & &&}
\startdata
11.0$\pm$0.2   &3.9 & 2.0 & 15.0 & 13.8 & 8.7 & 2.1 \\ 
11.35$\pm$0.15 & 4.7  & 1.5 & 21.2 & 8.4 & 9.1 & 1.6 \\
11.65$\pm$0.15 & 6.5  & 1.8 & 31.7 & 13.2  & 11.5 & 2.2 \\
12.0$\pm$0.2   &4.4   & 2.3 & 8.5 & 8.1  & 5.0 & 2.1 \\
\enddata
\end{deluxetable}

\subsection{Source Confusion}

The angular resolutions of the MIPS 24$\mu m$ maps and 
GALEX NUV maps are very well matched, both having FWHM$\sim 6"$.
The astrometry of GALEX sources is accurate to $\sim 1"$
(Seibert et al. 2005), and that of SWIRE sources is even better
(Surace et al. 2004). This ensures minimal mismatches between sources
in the two bands. The contamination from nearby foreground
bright sources is insignificant in the measurement of the 24$\mu m$ fluxes
of the NUV sources. According to published 24$\mu m$
number counts (Papovich et al. 2004;
Shupe et al. 2006), the chance for a random 
source of f$_{24}>0.2$ mJy to fall into the beam of an NUV
source is less than 1\%. For the measurement of the NUV fluxes
of 24$\mu m$ sources, the chance of this contamination is
higher (at $\sim 2\%$ level according to the NUV counts of
Xu et al. 2005). In order to minimize it, we did the stacking of the
NUV images after subtracting sources brighter
than NUV=23.5 mag (Section 3.1). Our choice of using the 'trimmed mean'
to estimate the average fluxes in the luminosity bins should
exclude sources seriously affected 
by bright neighbors. The confusion due to fainter sources adds another
noise (confusion noise) to the total error budget. It could biases
the measured flux of a stacked image to higher value only 
if the source population is strongly clustered (Takeuchi \& Ishii 2004). 
For the NUV
and 24$\mu m$ sources, this is not the case. Heinis et al.
(2004) found that UV galaxies are only very weakly clustered.
For the 24$\mu m$ sources, as shown by Zheng et al. (2006),
the confusion noise behaves very close to the random Gaussian noise down
to very faint flux level ($\sim 0.3$-- 0.4$\mu$Jy).
Uncorrelated
confused sources add an uniform diffuse background on the image,
which is subtracted in the normal background subtraction task
during the flux measurement. In summary, uncertainties due
to source confusion are unlikely to introduce significant
bias to our results.

\section{Discussion}

\subsection{Difference between z=0.6 and z=0 LIRGs}
In their study of faint galaxies in GOODS-N field, Melbourne et al. (2005) 
concluded that there is 
strong evidence for a morphological evolution of
the populations of LIRGs since redshift z=1. They found that 
above z=0.5, roughly half of all LIRGs are spirals and the 
peculiar/irregular-to-spiral ratio is $\sim 0.7$, whereas at low z,
spirals count for only one-third of LIRGs and the peculiar/irregular-to-spiral
ratio is 1.3. Similarly, Bell et al. (2005) found that at z$\sim 0.7$,
the IR luminosity density for $\rm 10.7 \la \log(L_{IR}/L_\sun) \la 11.5$
is dominated by spiral galaxies, and the contribution from clearly
interacting galaxies with morphology suggestive of major mergers is at most
30\%. Our result of the significantly lower mean L$_{dust}$/L$_{FUV}$ ratios (i.e. 
dust attenuation) for the z=0.6 
LIRGs compared to those of their local counterparts, is in
line with these findings. As shown in local samples (Sanders \& Mirabel
1996), non-interacting LIRGs usually are large, gas-rich
spirals with widely distributed
star formation all over the disk. In contrast, most of major-merger LIRGs 
have their starformation concentrates in the nuclei. These
nuclear starbursts are highly compact, and they generally show very high
dust attenuation (and warmer IR colors). If indeed the composite of LIRGs
has changed from spiral dominant at z$\sim 0.6$ to major-merger 
dominant at z=0, the increase of the mean dust attenuation for
these galaxies since z=0.6, as revealed in this work, is expected.
On the other hand, even at high z, the population of ULIRGs
is still dominated by major-mergers (Bell et al. 2005). This provides
a simple explanation on why the difference between the mean
L$_{dust}$/L$_{FUV}$ ratios of z=0.6 and z=0 galaxies does not
extend to higher luminosity bins in Fig.3.

\subsection{Evolution of cosmic dust attenuation}
There has been strong evidence for a positive (backward) evolution  
of the mean dust attenuation in star forming galaxies of z $\la 1$
(Takeuchi et al. 2005b).
From the GALEX deep survey of VVDS field,
Schiminovich et al. (2005) derived an evolution rate for
the UV luminosity density up to z=1 in the form of 
$\rho_{1500\AA} \propto (1+z)^{2.5\pm 0.7}$; at the same time,
Le Floc\'h et al. (2005) found from the Spitzer MIPS 24$\mu m$
deep survey of the CDFS field that the IR luminosity
function evolves between $0 \leq z \leq 1$ as
$\rm L^*_{IR} \propto (1+z)^{3.2^{+0.7}_{-0.2}}$ and
$\rm \phi^*_{IR} \propto (1+z)^{0.7^{+0.2}_{-0.6}}$,
corresponding to an IR luminosity density evolution of
$\rho_{IR} \propto (1+z)^{3.9}$. Therefore, the ratio
between the IR and UV luminosity densities increases
by a factor of $(1+0.6)^{1.4} = 1.9$ from z=0 to z=0.6,
indicating a significant increase of cosmic dust attenuation 
during this redshift interval.

Interestingly, in this work, it is found that for galaxies of
given UV or IR luminosities the dust attenuation did not
increase with the redshift. Actually, there is evidence
that the dust attenuation in z=0.6 LIRGs is even {\it lower}
than that in the local LIRGs. Is this consistent with
the positive backward evolution of cosmic dust attenuation? 
The key to understanding the apparent contradiction 
lies in the strong
dependence of the dust attenuation on the SFR
(Buat \& Burgarella 1998; Heckman et al. 1998; Martin et al. 2005;
Xu et al. 2006), and in 
the fact that SFR of the `average' starforming galaxy at z=0.6 
is much higher than the `average' starforming galaxy at z=0. 
It is worthwhile to check out whether this interpretation works quantitatively.
Assume the SFR of the average starforming galaxy evolving as
$(1+z)^3$, which is consistent with luminosity evolution of both UV and
IR galaxies. Then take the local (z=0) $\rm L_{dust}/L_{FUV}$ versus SFR relation,
which can be approximated by a linear dependence $\rm L_{dust}/L_{FUV} \propto SFR$
above SFR$\sim 0.1$ M$_\sun$ yr$^{-1}$ (Xu et al. 2006). Accordingly, 
from z=0 to z=0.6, 
the average $\rm L_{dust}/L_{FUV}$ ratio should increase by a factor 
of $(1+0.6)^3 = 4.1$, about a factor of 2 more than what is observed.
Hence, it appears that 
combining the cosmic SFR evolution with local $\rm L_{dust}/L_{FUV}$ 
versus SFR relation predicts {\it too much} evolution of cosmic dust
attenuation.
This again suggests that the major population of starforming galaxies
at z=0.6, of which contribution from LIRGs can be significant
(Hammer et al. 2005; Bell et al. 2005), may have less dust attenuation 
than that of z=0 galaxies of the same SFR.

\section{Conclusion} 

Using new SWIRE observations in the IR and GALEX observations in the UV, we
study the dust attenuation and stellar mass of two samples of z$\sim
0.6$ galaxies in the SWIRE/GALEX field ELAIS-N1-00 ($\Omega = 0.8$
deg$^2$).  The first sample is UV selected, having 600 galaxies with
photometric redshift $0.5 \leq z \leq 0.7$ and NUV$\leq 23.5$ mag
(corresponding to $\rm L_{FUV} \geq 10^{9.6}\; L_\sun$ for z=0.6).
The second sample is IR selected, containing 430 galaxies with
$f_{24\mu m} \geq 0.2$ mJy ($\rm L_{dust} \geq 10^{10.8}\; L_\sun$ at
z=0.6) and in the same photometric redshift range.  The $L_{dust}$ is
derived from the rest-frame 15$\mu$ luminosity.  The dust attenuation
is estimated using the luminosity ratio $\rm L_{dust}/L_{FUV}$.
Because of the low 24$\mu m$ detection rate (20\%) of the UV galaxies
and the low UV detection rate (27\%) of the IR galaxies, the stacking
technique is exploited in deriving mean $\rm L_{dust}/L_{FUV}$ ratios
in given $\rm L_{FUV}$ and $\rm L_{dust}$ bins for UV and IR selected
samples, respectively. The stellar mass is derived using the SWIRE
3.6$\mu m$ flux which measures the rest-frame K-band 2.2$\mu m$
emission. These results are compared to $\rm L_{dust}/L_{FUV}$ ratios
and the stellar mass of galaxies in control samples at z=0. It is
found that the mean $\rm L_{dust}/L_{FUV}$ ratios of the z=0.6 UV
galaxies are consistent with that of their z=0 counterparts of the
same $\rm L_{FUV}$. For IR galaxies, the mean $\rm L_{dust}/L_{FUV}$
ratios of the z=0.6 LIRGs are about a factor of 2 lower than local
LIRGs, whereas z=0.6 ULIRGs have the same mean $\rm L_{dust}/L_{FUV}$
ratios as their local counterparts. This is consistent with
results in the literature that show evidence of population changes 
of LIRGs from major-merger dominant at z=0 to spiral dominant
at z$> 0.5$. The stellar mass of z=0.6 UV
galaxies of $\rm L_{FUV} \leq 10^{10.2}\; L_\sun$ is about a factor 2
less than their local counterparts of the same luminosity, indicating
growth of these galaxies.  The mass of z=0.6 UVLGs ($\rm L_{FUV} >
10^{10.2}\; L_\sun$) and IR selected galaxies, which are nearly
exclusively LIRGs and ULIRGs, is the same as their local counterparts.

\vskip0.2truecm
\noindent{\it Acknowledgments}:

GALEX (Galaxy Evolution Explorer) is a NASA Small Explorer, launched
in April 2003.  We gratefully acknowledge NASA's support for
construction, operation, and science analysis for the GALEX mission,
developed in cooperation with the Centre National d'Etudes Spatiales
of France and the Korean Ministry of Science and Technology.  Support
for this work, part of the Spitzer Space Telescope Legacy Science
Program, was provided by NASA through an award issued by JPL under
NASA contract 1407. This publication makes use of data products from
the Two Micron All Sky Survey, which is a joint project of the
University of Massachusetts and the Infrared Processing and Analysis
Center/California Institute of Technology, funded by NASA and NSF.


\end{document}